\documentclass{ws-ijmpa}

\usepackage[super]{cite}
\usepackage{xcolor}
\usepackage[verbose,hypertexnames=false]{hyperref}
\hypersetup{colorlinks=false,allbordercolors=blue,pdfborderstyle={/S/U/W 1}}
\usepackage{lineno}

\begin{document}

\markboth{Balazs Korodi for the STAR Collaboration}{Measurement of spectator fragment pseudorapidity distributions with the STAR EPD}

%
\catchline{}{}{}{}{}
%

\title{Measurement of spectator fragment pseudorapidity distributions with the STAR Event Plane Detector
}

\author{Balazs Korodi for the STAR Collaboration}

\address{The Ohio State University, 281 W. Lane Ave\\
Columbus, 43210, USA}

\maketitle

\begin{history}
\received{Day Month Year}
\revised{Day Month Year}
\accepted{Day Month Year}
\published{Day Month Year}
\end{history}

\begin{abstract}
At sufficiently low center-of-mass energies, charged particles originating from the spectator region fall within the pseudorapidity acceptance ($2.14<|\eta|<5.09$) of the STAR Event Plane Detector (EPD). Measurements of the pseudorapidity distributions of helium spectator fragments are presented for Au+Au collisions at center-of-mass energies per nucleon pair of $\sqrt{s_{_{\mathrm{NN}}}}=$ 19.6, 14.6, 11.5, 9.2, and 7.7 GeV. The helium fragment yields in each EPD ring are extracted from fits to the deposited energy spectra using convoluted Landau distributions while accounting for the contribution from singly charged particles. The effects of the EPD detection efficiency and interactions of the fragments with the STAR beam pipe are evaluated using Geant4 Monte Carlo simulations. The resulting pseudorapidity distributions are presented as a function of collision energy and collision centrality.
\end{abstract}


\section{Introduction}	

In relativistic heavy-ion collisions, the fraction of nucleons participating in the primary inelastic interaction depends on the collision geometry. Nucleons that undergo at least one inelastic collision with nucleons from the opposing nucleus are referred to as participants, while those that avoid such interactions are known as spectators~\cite{Bialas:1976ed}. The participant matter forms the hot and dense medium created in the collision and is responsible for producing the majority of the observed final-state particles. In contrast, spectator matter remains largely outside the interaction zone and continues approximately along the original beam direction. Depending on the collision geometry and the nuclear breakup process, the spectator system may emerge as individual nucleons or as nuclear fragments of various masses, ranging from light nuclei such as helium and lithium to heavier residual fragments~\cite{PHOBOS:2015aeh}. Although spectators do not participate directly in the inelastic reaction, they can still undergo secondary processes, including elastic scattering, electromagnetic interactions, or fragmentation driven by the excitation of the remaining nuclear remnant. Consequently, the properties and distributions of spectator fragments provide valuable information about the collision geometry, the breakup of the projectile and target remnants, and the dynamics governing the spectator region.

One of the experimental facilities where spectator fragments can be investigated is the Relativistic Heavy Ion Collider (RHIC)~\cite{Chen:2024aom}. During the second phase of the Beam Energy Scan program (BES-II), conducted between 2018 and 2021, Au+Au collisions were recorded over a center-of-mass energy per nucleon pair range of $\sqrt{s_{_{\mathrm{NN}}}}=27$--$7.7$ GeV in collider mode. At these collision energies, spectator fragments are emitted at sufficiently small angles with respect to the beam axis that they bypass the central detectors, while a significant fraction of the charged spectator fragments remains within the acceptance of dedicated forward detectors. One such detector is the Event Plane Detector (EPD)~\cite{Adams:2019fpo}, which was installed in 2018 as part of the Solenoidal Tracker at RHIC (STAR) experiment. Covering a broad forward and backward pseudorapidity range, the EPD is capable of detecting charged particles, including spectator fragments, making it well suited for measurements of spectator fragment pseudorapidity distributions.

In this paper, measurements of the pseudorapidity distributions of helium spectator fragments in Au+Au collisions at $\sqrt{s_{_{\mathrm{NN}}}}=19.6$, 14.6, 11.5, 9.2, and 7.7 GeV are presented. The helium fragment yields are extracted from fits to the EPD energy-loss spectra using convoluted Landau functions and are corrected for the detector acceptance and reconstruction efficiency using Monte Carlo (MC) simulations. The resulting pseudorapidity distributions are presented for a range of collision centralities and beam energies, and their dependence on these collision parameters is discussed.

\section{Experimental setup}

A general description of the STAR detector is provided in Ref.~\cite{STAR:2002eio}. In the present analysis, only the EPD~\cite{Adams:2019fpo} is utilized (in addition to the detectors used for event selection and categorization).

\begin{figure}[h]
\centering
\includegraphics[width=0.6\linewidth,trim={2cm 0 6cm 0},clip]{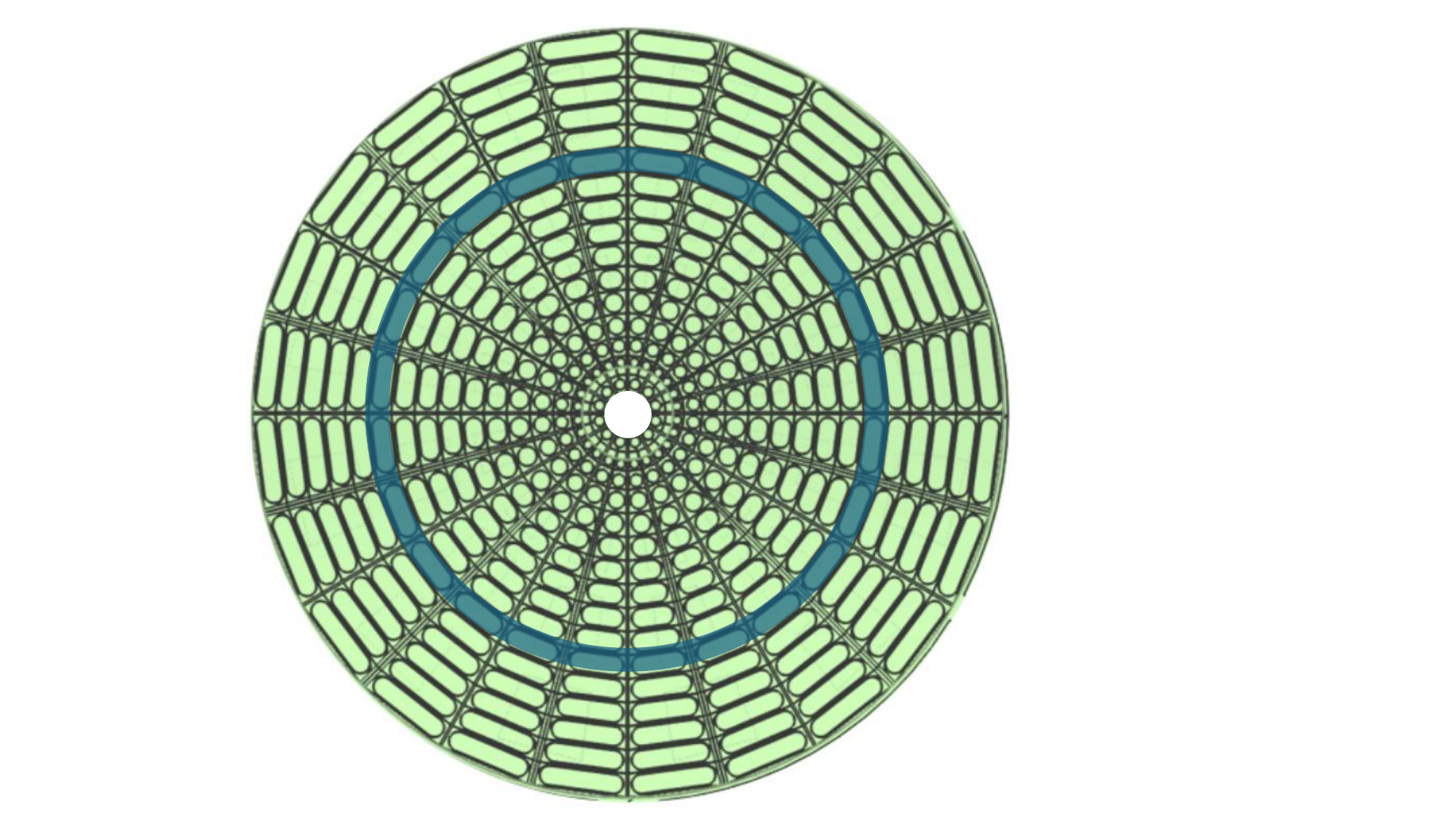}
\caption{Schematic illustration of an EPD wheel, with one of the 16 rings highlighted in blue.}
\label{f:epdstructure}
\end{figure}

The EPD consists of two highly segmented scintillator wheels, positioned symmetrically at $\pm375$ cm from the nominal interaction point along the beam axis and oriented perpendicular to it. This geometry provides acceptance for charged particles over the pseudorapidity interval $2.14<|\eta|<5.09$, where pseudorapidity is defined as $\eta=-\ln[\tan(\theta/2)]$, with $\theta$ denoting the polar angle with respect to the beam axis. Each wheel is segmented into 16 concentric rings, each corresponding to a distinct pseudorapidity interval listed in Table~\ref{t:eta_multicol}. Every ring is composed of 24 scintillator tiles, except for the innermost ring, which contains 12 tiles. A schematic view of an EPD wheel is shown in Fig.~\ref{f:epdstructure}. Scintillation light produced in the tiles is collected by wavelength-shifting fibers and transported to silicon photomultipliers (SiPMs) mounted on the STAR magnet pole tips. The resulting signals are digitized by the STAR data acquisition system. Since the EPD measures only the ionization energy deposited by charged particles, it provides only limited particle identification capabilities that will be highlighted later.

\begin{table}[h]
\tbl{Columns 2--5 summarize the geometrical acceptance of the EPD rings. Columns 2 and 3 list the minimum ($r_i$) and maximum ($r_f$) radial distances from the beam axis, while Columns 4 and 5 give the corresponding maximum ($\eta_i$) and minimum ($\eta_f$) pseudorapidity values for particles emitted from the nominal interaction point and traveling in straight lines~\cite{Adams:2019fpo}.\label{t:eta_multicol}}
{\begin{tabular}{@{}ccccc@{}} \toprule
\textbf{Ring} & $r_i$ (cm) & $r_f$ (cm) & $\eta_i$ & $\eta_f$ \\
\colrule
0  & 4.6    & 9.0    & 5.09 & 4.42 \\
1  & 9.0    & 13.4   & 4.42 & 4.03 \\
2  & 13.4   & 17.8   & 4.03 & 3.74 \\
3  & 17.8   & 23.3  & 3.74 & 3.47 \\
4  & 23.3  & 28.9  & 3.47 & 3.26 \\
5  & 28.9  & 34.4  & 3.26 & 3.08 \\
6  & 34.4  & 39.9  & 3.08 & 2.94 \\
7  & 39.9  & 45.5  & 2.94 & 2.81 \\
8  & 45.5  & 51.0  & 2.81 & 2.69 \\
9  & 51.0  & 56.5  & 2.69 & 2.59 \\
10 & 56.5  & 62.1  & 2.59 & 2.50 \\
11 & 62.1  & 67.6  & 2.50 & 2.41 \\
12 & 67.6  & 73.1  & 2.41 & 2.34 \\
13 & 73.1  & 78.6  & 2.34 & 2.27 \\
14 & 78.6  & 84.2  & 2.27 & 2.20 \\
15 & 84.2  & 89.7  & 2.20 & 2.14 \\
\botrule
\end{tabular}}
\end{table}

\section{Analysis details}

\subsection{Data sets}

The analysis is based on minimum-bias Au+Au collision data collected with the STAR detector at $\sqrt{s_{_{\mathrm{NN}}}}=19.6$, 14.6, 11.5, 9.2, and 7.7 GeV during the 2019--2021 RHIC running periods. Selected events were required to contain a primary collision vertex reconstructed with the Time Projection Chamber (TPC) within $\pm15$ cm of the nominal interaction point along the beam direction. In addition, the transverse displacement of the primary vertex from the beam axis was required to be smaller than 1 cm.

The results are presented for the 5--10\%, 10--20\%, 20--30\%, 30--40\%, 40--50\%, 50--60\%, and 60--70\% centrality intervals. Collision centrality was determined from the charged-particle multiplicity measured by the TPC within $|\eta|<0.5$. The multiplicity distributions were corrected for luminosity and vertex-dependent acceptance effects before being compared with Monte Carlo Glauber calculations to define the centrality classes. A detailed description of the centrality determination procedure is available in Ref.~\cite{STAR:2012och}.

Since the acceptance of the EPD does not overlap with that of the STAR tracking detectors, the analysis relies exclusively on the EPD response after the event selection criteria are applied. 

\subsection{Landau fits}

The EPD measures the ionization energy deposited by charged particles in its scintillator tiles, with the signal recorded in units of Analog-to-Digital Converter (ADC) counts. Owing to the stochastic nature of energy loss, particles with identical kinematic properties do not deposit a fixed amount of energy. For minimum ionizing particles (MIPs) traversing a thin scintillator, the energy-loss distribution follows a Landau distribution~\cite{Adams:2019fpo}, characterized by its most probable value (MPV) and width (WID). Following data taking, the recorded ADC spectrum of each EPD tile is calibrated by applying a scale factor such that the MPV of the single-MIP distribution is normalized to unity.

When multiple charged particles traverse the same tile, the measured energy-loss distribution is given by the convolution of the corresponding number of single-particle Landau distributions. Since the number of particles hitting a tile varies from event to event, the measured ADC spectrum is modeled as a weighted sum of convoluted Landau distributions,
\begin{equation} \label{e:Landaufit}
    \frac{dN(x)}{d(\text{Calibrated ADC})}=\sum_{n=1}^{n_{\text{max}}}M_nL_n(x),
\end{equation} 
where $n_{\text{max}}$ is the largest number of particles considered hitting the tile, each $M_n$ is a parameter (also called n-MIP weight) representing the probability of $n$ MIPs hitting the tile, and $L_n$ are the convoluted Landau distributions:
\begin{equation}
    L_n(x)=\begin{cases} \text{Landau(x; MPV, WID)}, & \text{if} \hspace{0.2cm} n=1; \\
                     \int_{-\infty}^{\infty} L_{n-1}(x-y) L_1(y)dy, & \text{otherwise},
       \end{cases}
\end{equation}
where $x$ and $y$ are the variables of the distribution.

\begin{figure}[h]
\centerline{\includegraphics[width=9 cm]{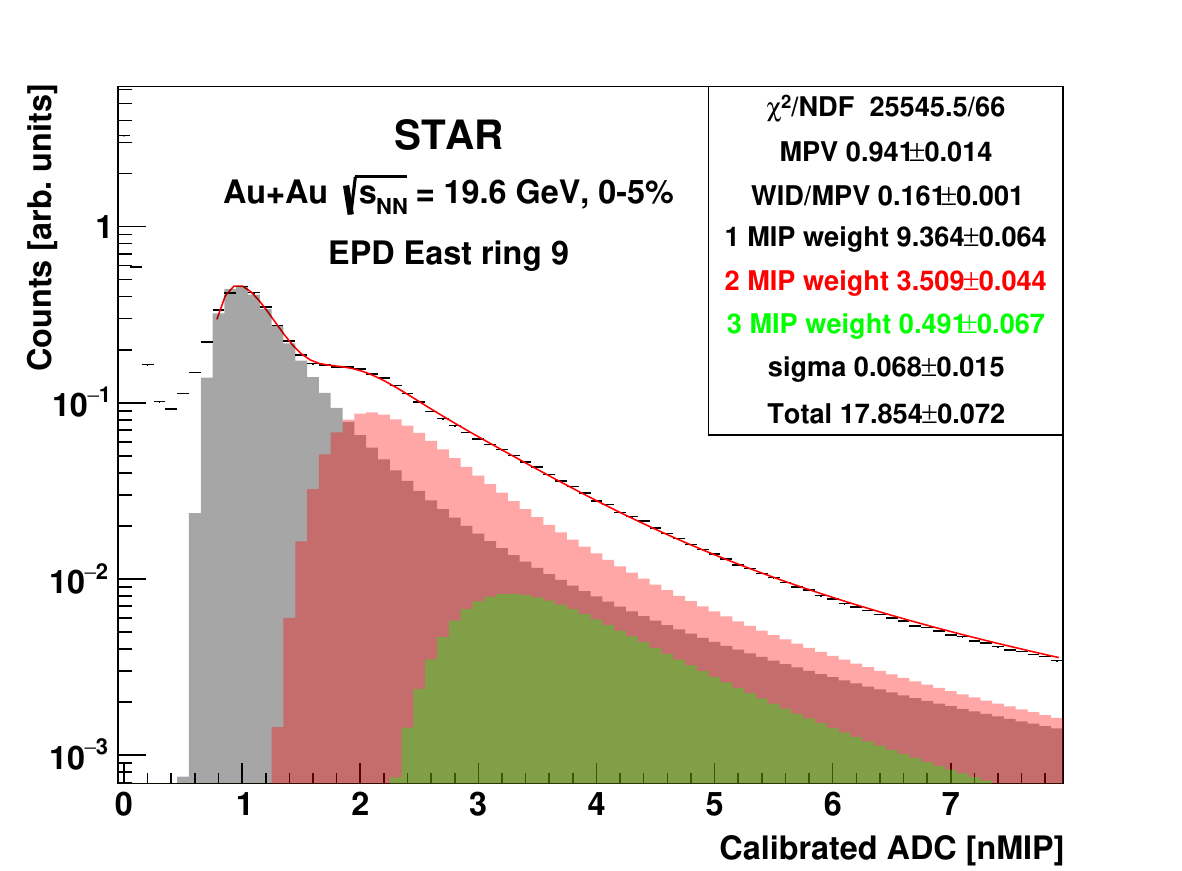}}
\caption{Convoluted Landau fit to the calibrated ADC spectrum of an EPD ring. The data are shown as black markers, while the red curve represents the fit up to a calibrated ADC value of 8. The colored regions indicate the contributions from 1-, 2-, and 3-MIP components. The fitted parameter values are listed in the legend. The average number of MIPs contributing to the spectrum is given as the ``Total''.}
\label{f:Langaufit}
\alttext{Landau fit at 19.6 GeV}
\end{figure}

In experimental data, additional detector and physics effects, including the angular distribution of the incident particles, deviations from minimum-ionizing behavior, and electronic noise, broaden the measured ADC spectrum. This broadening is incorporated by convolving the Landau model with a Gaussian distribution,
\begin{equation} \label{e:Langaufit}
    \frac{dN(x)}{d(\text{Calibrated ADC})}=\sum_{n=1}^{n_{\text{max}}}M_n\int_{-\infty}^\infty  L_n(x-y)G(y)dy,
\end{equation} 
where $G(x)$ is a normalized Gaussian distribution. An example fit to a calibrated ADC spectrum using this parameterization is shown in Fig.~\ref{f:Langaufit} for $\sqrt{s_{_{\mathrm{NN}}}}=19.6$ GeV. Given the large statistical precision of the data and the approximate nature of the model used to describe the ADC spectra, the fits are not expected to yield $\chi^2/\mathrm{NDF}\approx 1$, where NDF denotes the number of degrees of freedom.

\subsection{Extraction of fragment yields}

The description of the EPD spectra presented above is valid only when all incident particles are singly charged. At lower collision energies, however, spectator fragments with different charge states fall within the acceptance of the EPD. Since these fragments originate from the spectator remnants, they retain a large fraction of the beam momentum and therefore behave approximately as minimum ionizing particles. Consequently, they appear in the ADC spectra of the inner EPD rings in non-central collisions as additional Landau peaks, with the most probable value located at a calibrated ADC value of approximately $Q^2$, where $Q$ is the charge of the fragment.

 Assuming that the incident singly charged particles are uncorrelated, their $n$-MIP weights are expected to follow a Poisson distribution. To distinguish the fragment contribution from higher-order singly charged $n$-MIP contributions, this Poisson behavior is imposed as a constraint during the fit. As a result, the $n$ independent $n$-MIP weight parameters are replaced by a single free parameter corresponding to the mean of the Poisson distribution.

\begin{figure}[h]
\centerline{\includegraphics[width=9 cm]{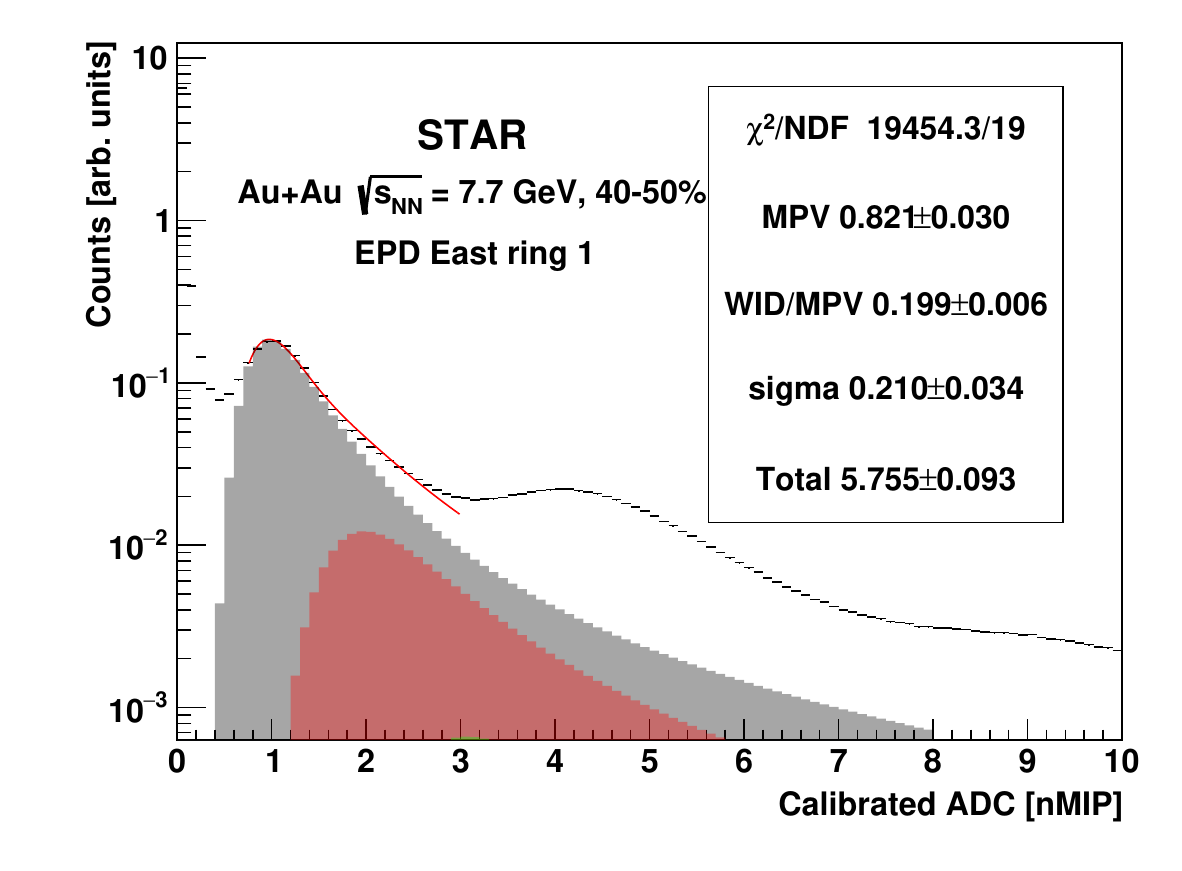}}
\caption{Convoluted Landau fit with the Poisson constraint applied to the singly charged particle-dominated region of the calibrated ADC spectrum of an EPD ring. The data are shown as black markers, while the red curve represents the fit up to a calibrated ADC value of 3. The colored regions indicate the contributions from the 1- and 2-MIP components. The fitted parameter values are listed in the legend. The average number of singly charged MIPs contributing to the spectrum is given as the ``Total''.}
\label{f:LangaufitMIP}
\alttext{Landau fit to the singly-charged dominated part}
\end{figure}

\begin{figure}[h]
\centerline{\includegraphics[width=9 cm]{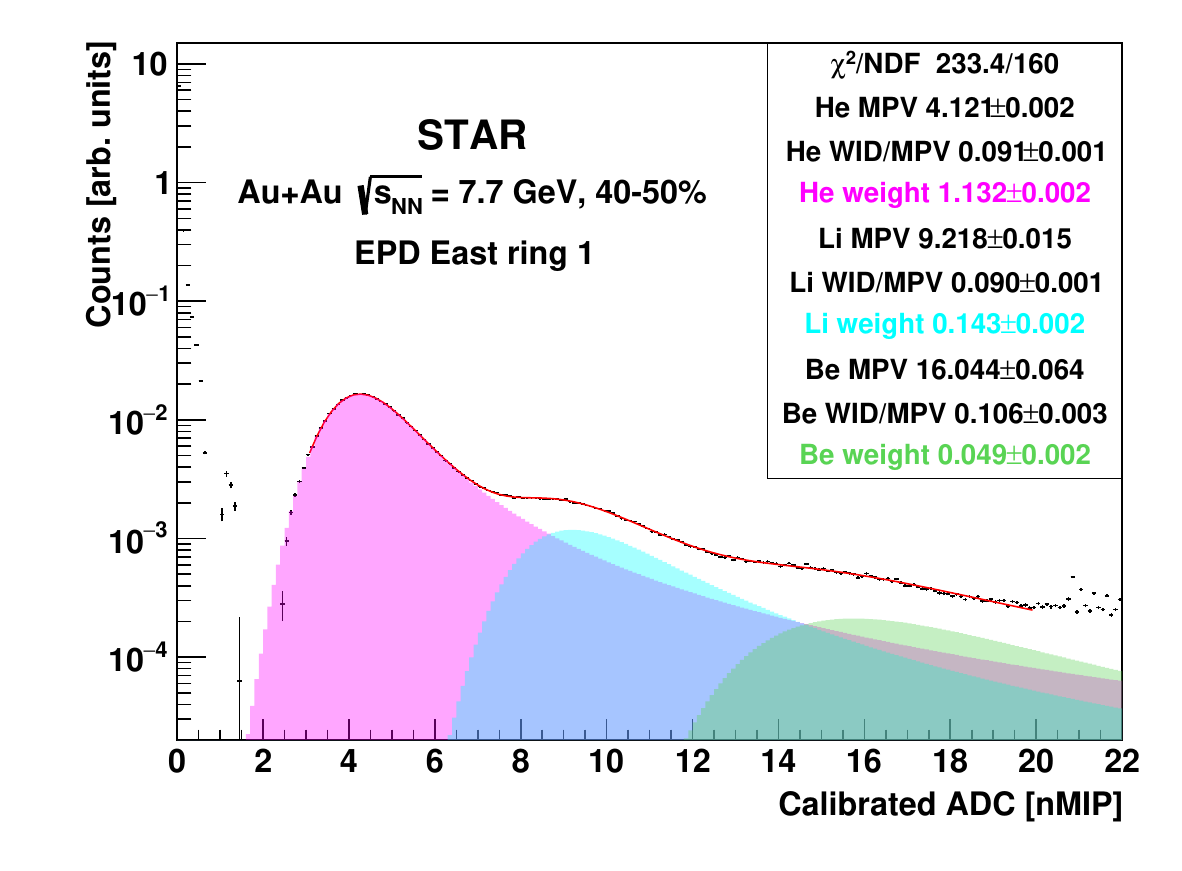}}
\caption{Landau fit to the calibrated ADC spectrum of spectator fragments in a given EPD ring. The data are shown as black markers, while the red curve represents the fit. The colored regions indicate the contributions from helium, lithium, and beryllium fragments. The fitted parameter values are listed in the legend.}
\label{f:Langaufitalpha}
\alttext{Landau fit to the fragment part}
\end{figure}

 To extract the spectator fragment yields, the contribution from singly charged particles is first determined and subtracted. The singly charged particle-dominated region of the calibrated ADC spectrum (calibrated ADC $<3$) is fitted using Eq.~\ref{e:Langaufit} with the Poisson constraint applied. An example of such a fit is shown in Fig.~\ref{f:LangaufitMIP}. The fitted singly charged contribution is then subtracted from the measured spectrum. The remaining fragment spectrum is described by
\begin{equation} \label{e:fragmentfit}
    \frac{dN_{\text{fragment}}(x)}{d(\text{Calibrated ADC})}=M_{\text{He}}L_{\text{He}}(x) + M_{\text{Li}}L_{\text{Li}}(x) + M_{\text{Be}}L_{\text{Be}}(x),
\end{equation} 
where $M_{\mathrm{He}}$, $M_{\mathrm{Li}}$, and $M_{\mathrm{Be}}$ are the fragment yields (weights), while $L_{\mathrm{He}}$, $L_{\mathrm{Li}}$, and $L_{\mathrm{Be}}$ denote the corresponding Landau distributions for helium, lithium, and beryllium fragments, respectively. An example fit to the fragment spectrum is shown in Fig.~\ref{f:Langaufitalpha}. Separate fits are done for each collision energy, centrality class, and EPD ring.

\subsection{Efficiency correction}

The detection efficiency for spectator fragments is less than 100\%. The intrinsic efficiency of the EPD is approximately 95\% for the inner rings due to the gaps between adjacent scintillator tiles. In addition, spectator fragments may undergo nuclear interactions while traversing the beam pipe, resulting in a further reduction of the detection efficiency.

The overall detection efficiency for helium fragments is determined using Geant4 simulations~\cite{Allison:2016lfl}. Alpha particles are generated with a uniform distribution in $\eta$ and propagated through the detector geometry. Although the EPD cannot distinguish between $^3$He and $^4$He, the helium spectator yield is expected to be dominated by alpha particles~\cite{Ogul:2010si}. The momentum of each simulated alpha particle is calculated by assuming that it travels with the beam velocity of the incoming nucleus. In this momentum range, the detection efficiency is found to be independent of the particle momentum. The particles are generated at the center of the TPC, and nearly all particles reaching the EPD follow straight trajectories and are detected in the ring corresponding to their generated pseudorapidity. For each EPD ring, the detection efficiency is calculated as the ratio of the number of alpha particles depositing more than 3 calibrated ADC in that ring to the number of alpha particles generated in the pseudorapidity interval covered by the ring. The efficiency is approximately 60\% for the innermost ring and increases toward larger radii, reaching about 80\% for the fifth ring.

The pseudorapidity distribution of helium fragments, $dN_{\mathrm{He}}/d\eta$, is obtained by correcting the helium yields extracted from the Landau fits for the detection efficiency and dividing by the width of the pseudorapidity interval corresponding to each EPD ring. The determination of the detection efficiencies for lithium and beryllium fragments is still in progress. Consequently, only the efficiency-corrected helium fragment results are presented in this paper.

\subsection{Systematic uncertainties}

The systematic uncertainties were evaluated by varying the centrality selection, comparing the results obtained with the East and West EPD wheels, and modifying the Landau fitting procedure. The dominant contribution was found to originate from the uncertainty in the Landau fitting procedure, which was quantified by varying the fitting approach, including modifications to the singly charged particle background, variations of the WID and Gaussian width parameters, and fitting the full ADC spectrum with a single fit. Among these studies, the treatment of the singly charged background was found to have the largest impact on the extracted helium yields.

The total systematic uncertainty was determined for each pseudorapidity bin by combining the contributions from the individual sources in quadrature, assuming them to be uncorrelated. The average systematic uncertainty for each collision energy and centrality class is shown in Table~\ref{tab:relative_systematic_uncertainties}.

\begin{table}[htbp]
\centering
\setlength{\tabcolsep}{4pt}
\renewcommand{\arraystretch}{1.3}
\caption{Average relative systematic uncertainties (\%). Upper and lower uncertainties are shown as superscripts and subscripts.}
\label{tab:relative_systematic_uncertainties}
\vspace{6pt}
\begin{tabular}{@{}cccccc@{}}
\hline
Centrality & \multicolumn{5}{c}{$\sqrt{s_{\mathrm{NN}}}$ (GeV)} \\
\cline{2-6}
 & 7.7 & 9.2 & 11.5 & 14.6 & 19.6 \\
\hline
5--10\%  & ${}^{+12.43}_{-18.13}$ & --- & --- & --- & --- \\
10--20\% & ${}^{+8.25}_{-10.84}$ & ${}^{+8.11}_{-11.10}$ & ${}^{+7.14}_{-11.08}$ & --- & --- \\
20--30\% & ${}^{+7.18}_{-8.41}$ & ${}^{+8.97}_{-9.84}$ & ${}^{+7.02}_{-11.64}$ & ${}^{+7.16}_{-19.31}$ & ${}^{+8.55}_{-34.59}$ \\
30--40\% & ${}^{+9.17}_{-7.97}$ & ${}^{+12.16}_{-9.18}$ & ${}^{+9.23}_{-11.45}$ & ${}^{+11.04}_{-18.75}$ & ${}^{+12.38}_{-29.83}$ \\
40--50\% & ${}^{+12.31}_{-8.29}$ & ${}^{+15.56}_{-9.36}$ & ${}^{+10.94}_{-11.31}$ & ${}^{+13.00}_{-17.07}$ & ${}^{+12.78}_{-23.61}$ \\
50--60\% & ${}^{+15.33}_{-8.68}$ & ${}^{+19.60}_{-9.56}$ & ${}^{+13.63}_{-11.10}$ & ${}^{+14.60}_{-15.80}$ & ${}^{+12.72}_{-20.00}$ \\
60--70\% & ${}^{+18.41}_{-9.07}$ & ${}^{+22.34}_{-9.09}$ & ${}^{+20.54}_{-10.84}$ & ${}^{+17.47}_{-14.41}$ & ${}^{+14.90}_{-16.62}$ \\
\hline
\end{tabular}
\end{table}

\section{Results and discussion}

\begin{figure}[h!]
\centering
\includegraphics[width=0.47\linewidth]{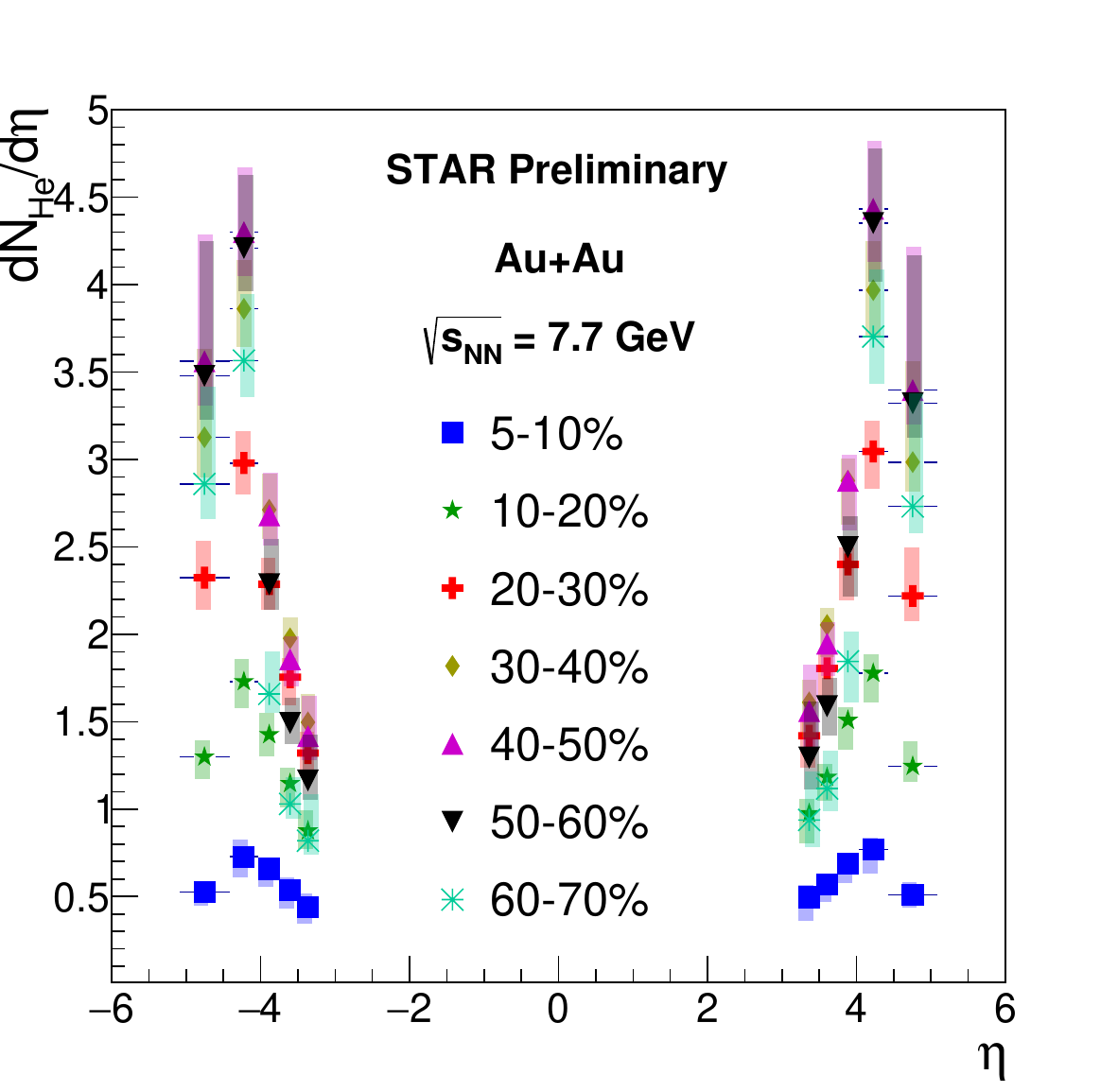}
\includegraphics[width=0.47\linewidth]{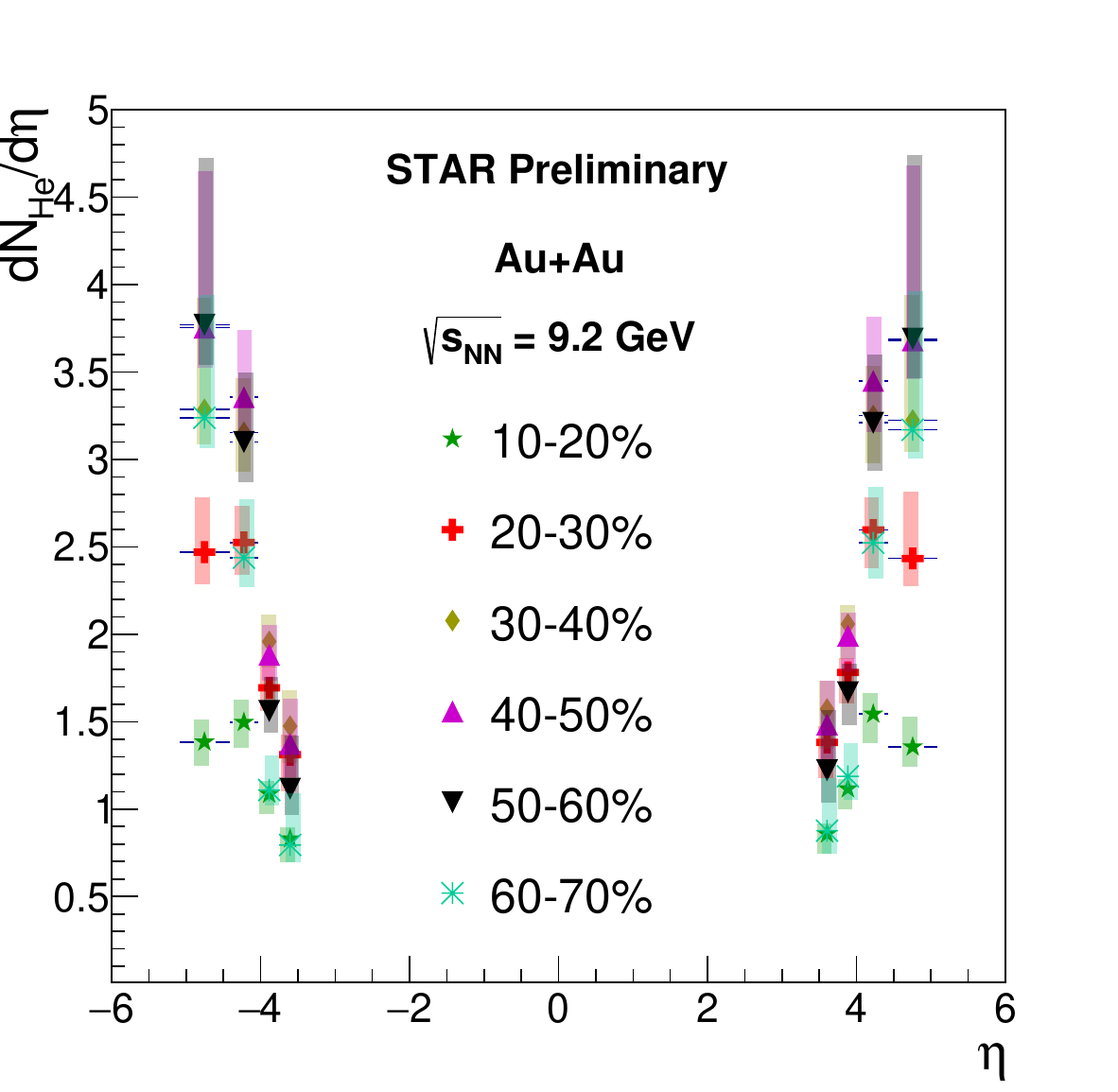}
\includegraphics[width=0.47\linewidth]{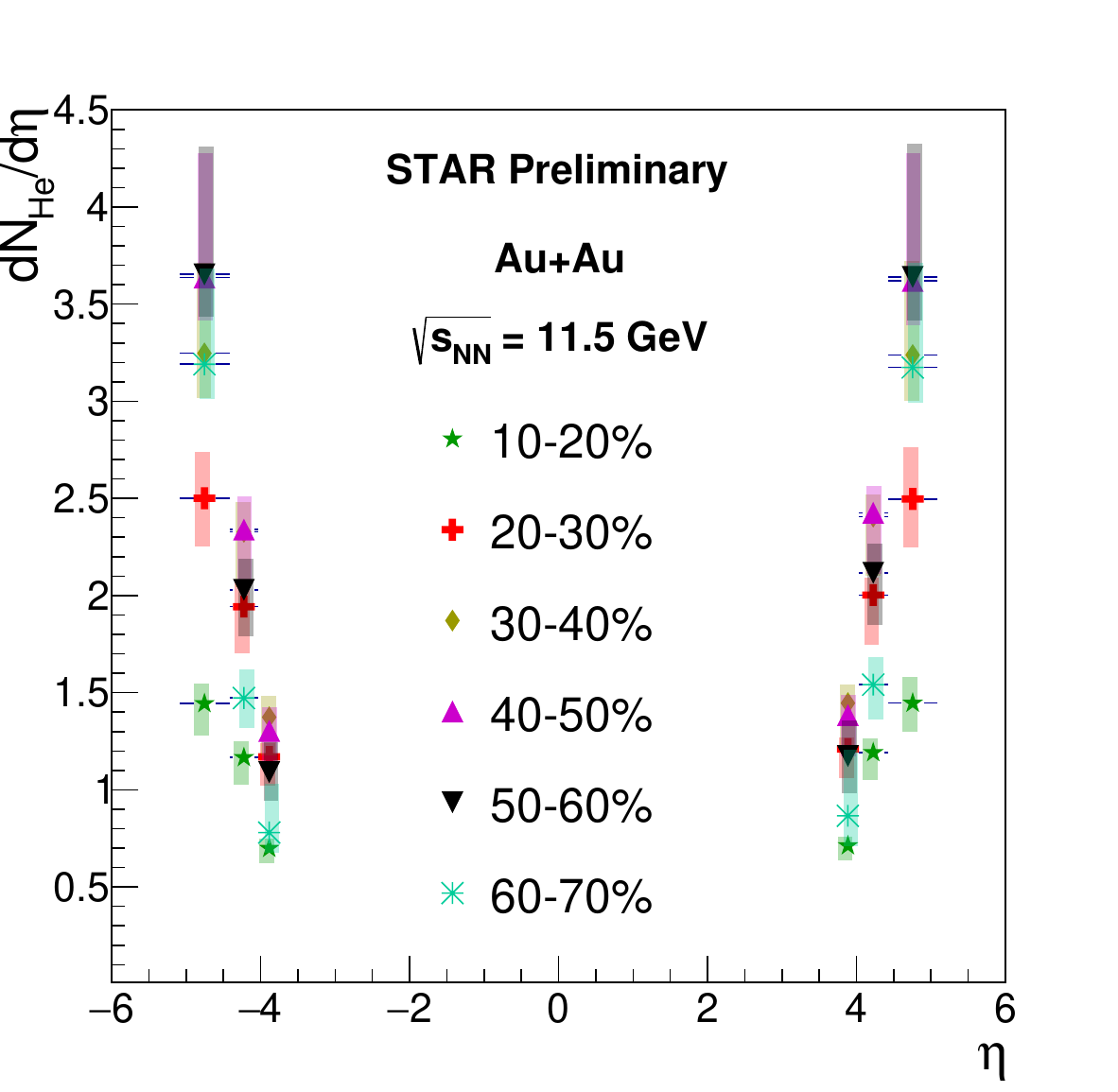}
\includegraphics[width=0.47\linewidth]{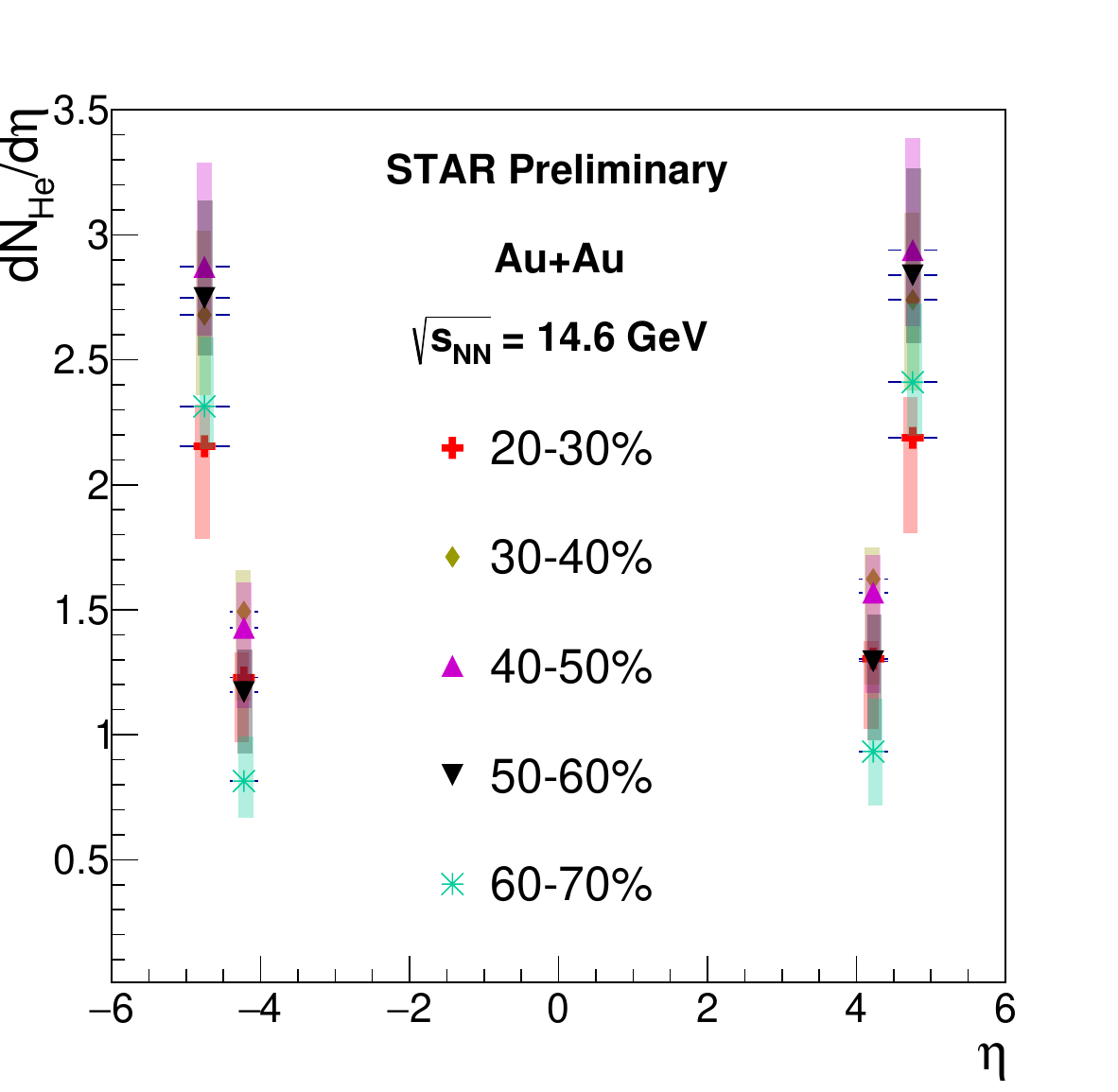}
\includegraphics[width=0.47\linewidth]{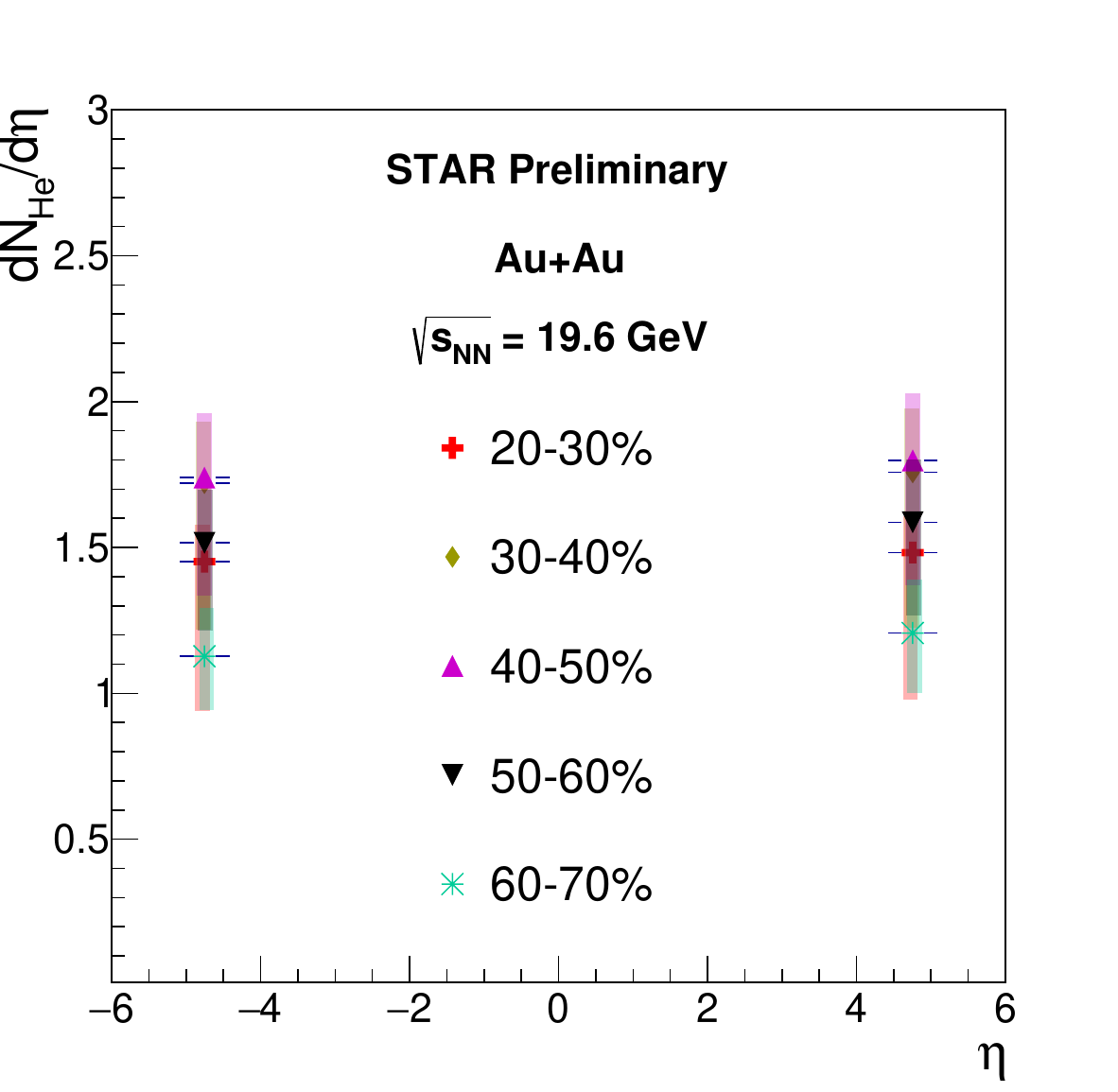}
\caption{Pseudorapidity distributions of helium fragments, $dN_{\mathrm{He}}/d\eta$, measured with the EPD at (from left to right, top to bottom) $\sqrt{s_{_{\mathrm{NN}}}}=$ 7.7, 9.2, 11.5, 14.6, and 19.6 GeV for different collision centralities. The shaded bands represent the systematic uncertainties. Statistical uncertainties are shown as error bars, but are generally smaller than the marker size.}
\label{f:helium_dNdeta}
\alttext{Helium fragment dN/deta}
\end{figure}

The measured pseudorapidity distributions of helium fragments are shown in Fig.~\ref{f:helium_dNdeta} for different collision energies and centrality classes. For a given collision energy, results are shown only for EPD rings in which a non-zero helium yield was extracted for all displayed centrality classes. The results at negative and positive pseudorapidity are obtained from independent measurements using the East and West EPD wheels, respectively. As the collision energy increases, the peak of the $dN_{\mathrm{He}}/d\eta$ distribution shifts toward larger values of $|\eta|$, resulting in non-zero yields being extracted from progressively fewer EPD rings. In addition, the increasing charged-particle multiplicity from the participant region at higher collision energies makes the extraction of non-zero fragment yields in the more central collisions increasingly difficult.

\begin{figure}[h]
\centering
\includegraphics[width=0.89\linewidth]{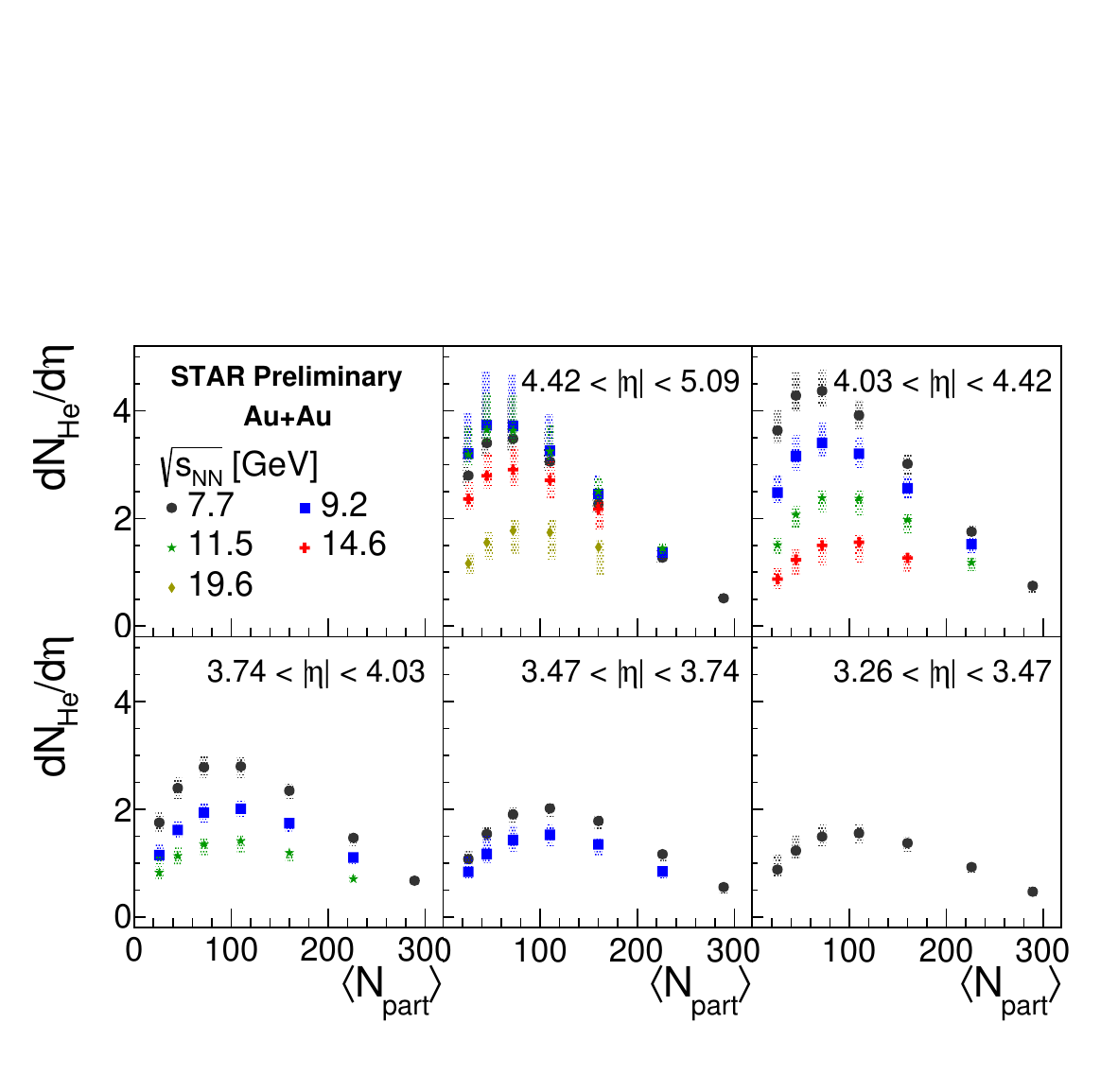}
\caption{Pseudorapidity distributions of helium fragments, $dN_{\mathrm{He}}/d\eta$, as a function of the average number of participants, $\langle N_{\mathrm{part}}\rangle$, for different collision energies. Each panel corresponds to a given EPD ring and shows the average of the East and West sides. The shaded bands represent the systematic uncertainties. Statistical uncertainties are shown as error bars, but are generally smaller than the marker size.}
\label{f:dNdetavsNpart}
\alttext{Helium fragment dN/deta versus the number of participants}
\end{figure}

To investigate the centrality dependence, Fig.~\ref{f:dNdetavsNpart} presents $dN_{\mathrm{He}}/d\eta$ as a function of the average number of participants, $\langle N_{\mathrm{part}}\rangle$, for each EPD ring and collision energy. The results shown are averaged over the East and West EPD wheels. For all collision energies, a non-monotonic dependence of $dN_{\mathrm{He}}/d\eta$ on $\langle N_{\mathrm{part}}\rangle$ is observed at fixed $|\eta|$, with a maximum occurring in the range $60<\langle N_{\mathrm{part}}\rangle<120$, depending on $|\eta|$. Within the acceptance of the EPD, the helium fragment yield increases with decreasing collision energy. This behavior is primarily a consequence of the shift of the pseudorapidity distributions toward larger values of $|\eta|$ with increasing collision energy; beyond the acceptance of the EPD, the opposite trend is expected.

\begin{figure}[h]
\centering
\includegraphics[width=0.89\linewidth]{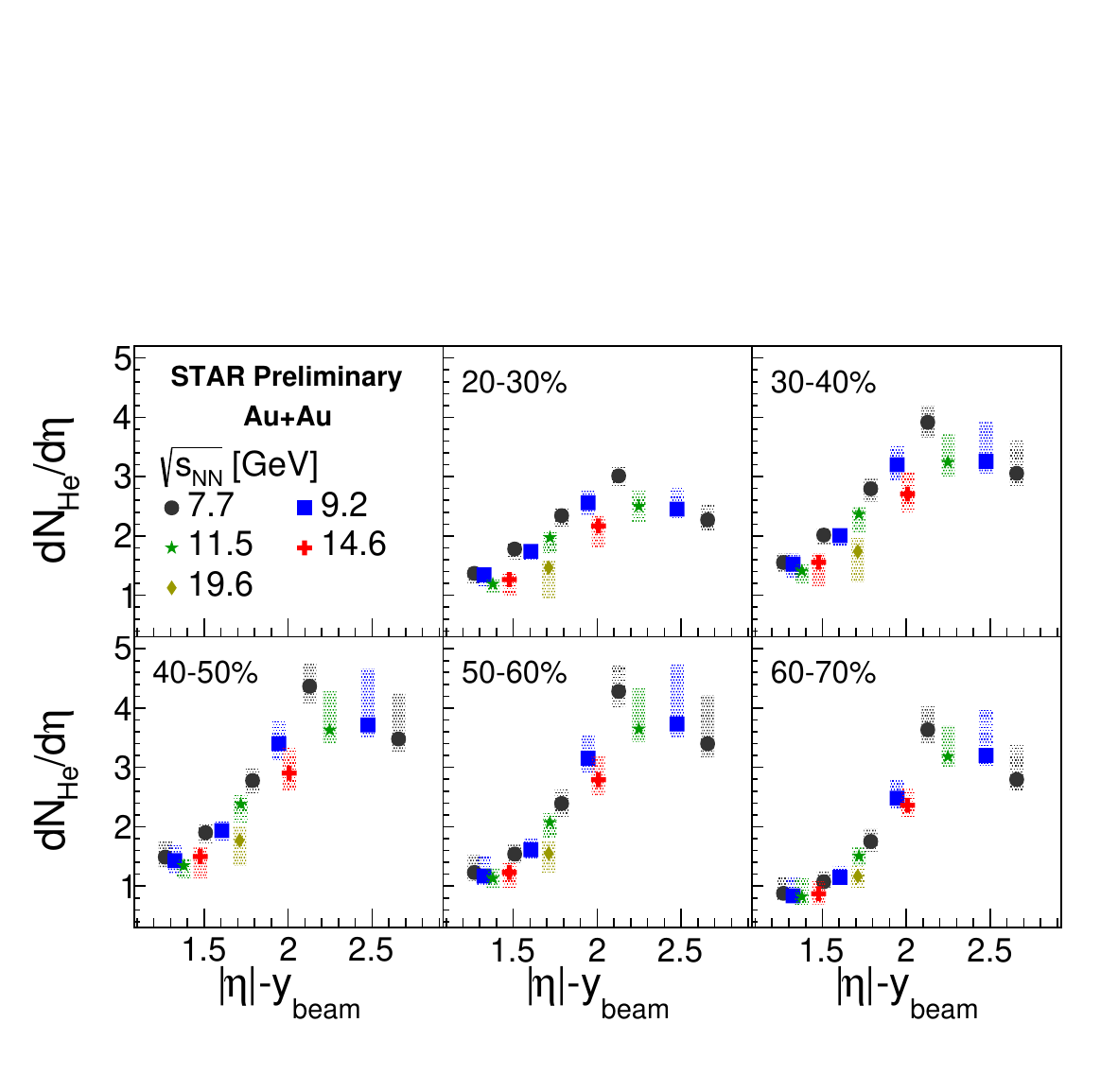}
\caption{Pseudorapidity distributions of helium fragments, $dN_{\mathrm{He}}/d\eta$, plotted as a function of $\eta-y_{\mathrm{beam}}$ for different collision energies. The shaded bands represent the systematic uncertainties. Statistical uncertainties are shown as error bars, but are generally smaller than the marker size.}
\label{f:dNdetalimiting}
\alttext{Helium fragment dN/deta shifted by beam rapidity}
\end{figure}

Figure~\ref{f:dNdetalimiting} shows $dN_{\mathrm{He}}/d\eta$ as a function of $\eta-y_{\mathrm{beam}}$, where $y_{\mathrm{beam}}$ denotes the beam rapidity. Only centrality classes for which measurements are available at all five collision energies are shown. The results from the East and West EPD wheels are again averaged. According to the limiting fragmentation hypothesis~\cite{Benecke:1969sh}, the distributions are expected to become independent of collision energy when expressed in terms of $\eta-y_{\mathrm{beam}}$. This expectation is qualitatively satisfied for the most peripheral collisions. In the more central collisions, however, a modest deviation from this behavior is observed, with $dN_{\mathrm{He}}/d\eta$ decreasing as the collision energy increases at fixed $\eta-y_{\mathrm{beam}}$. It should be noted that the limiting fragmentation hypothesis is, strictly speaking, formulated for rapidity rather than pseudorapidity distributions. Owing to the relatively large mass of helium fragments, the difference between rapidity and pseudorapidity is non-negligible and may contribute to the observed deviation from a universal limiting-fragmentation curve.

\section{Summary}

Measurements of helium fragment pseudorapidity distributions, $dN_{\mathrm{He}}/d\eta$, are presented for Au+Au collisions at $\sqrt{s_{_{\mathrm{NN}}}}=$ 19.6, 14.6, 11.5, 9.2, and 7.7 GeV using the STAR Event Plane Detector (EPD). Helium fragment yields are extracted from the deposited energy spectra using Landau fits and corrected for the detection efficiency based on Geant4 Monte Carlo simulations.

The $dN_{\mathrm{He}}/d\eta$ distributions peak within the EPD acceptance at the lower collision energies studied, while the peak shifts beyond the acceptance with increasing collision energy. The results exhibit a non-monotonic centrality dependence at all energies. A hint of a possible deviation from the universal behavior predicted by the limiting fragmentation hypothesis is observed.

\section*{Acknowledgments}

This work was supported by the U.S. Department of Energy, Office of Science, Office of Nuclear Physics, under Award No. DE-SC0020651.

\bibliographystyle{ws-ijmpa}
\bibliography{sample}

@article{STAR:2002eio,
    author = "Ackermann, K. H. and others",
    collaboration = "STAR",
    title = "{STAR detector overview}",
    doi = "10.1016/S0168-9002(02)01960-5",
    journal = "Nucl. Instrum. Meth. A",
    volume = "499",
    pages = "624--632",
    year = "2003"
}

@article{Adams:2019fpo,
    author = "Adams, Joseph and others",
    title = "{The STAR Event Plane Detector}",
    eprint = "1912.05243",
    archivePrefix = "arXiv",
    primaryClass = "physics.ins-det",
    doi = "10.1016/j.nima.2020.163970",
    journal = "Nucl. Instrum. Meth. A",
    volume = "968",
    pages = "163970",
    year = "2020"
}

@article{PHOBOS:2015aeh,
    author = "Back, B. B. and others",
    collaboration = "PHOBOS",
    title = "{Participant and spectator scaling of spectator fragments in Au+Au and Cu+Cu collisions at $\sqrt s_{NN}$ = 19.6 and 22.4 GeV}",
    eprint = "1511.07921",
    archivePrefix = "arXiv",
    primaryClass = "nucl-ex",
    doi = "10.1103/PhysRevC.94.024903",
    journal = "Phys. Rev. C",
    volume = "94",
    number = "2",
    pages = "024903",
    year = "2016"
}

@article{Bialas:1976ed,
    author = "Bialas, A. and Bleszynski, M. and Czyz, W.",
    title = "{Multiplicity Distributions in Nucleus-Nucleus Collisions at High-Energies}",
    reportNumber = "TPJU-9/76",
    doi = "10.1016/0550-3213(76)90329-1",
    journal = "Nucl. Phys. B",
    volume = "111",
    pages = "461--476",
    year = "1976"
}

@article{Chen:2024aom,
    author = "Chen, Jinhui and others",
    title = "{Properties of the QCD matter: review of selected results from the relativistic heavy ion collider beam energy scan (RHIC BES) program}",
    eprint = "2407.02935",
    archivePrefix = "arXiv",
    primaryClass = "nucl-ex",
    doi = "10.1007/s41365-024-01591-2",
    journal = "Nucl. Sci. Tech.",
    volume = "35",
    number = "12",
    pages = "214",
    year = "2024"
}

@article{STAR:2012och,
    author = "Adamczyk, L. and others",
    collaboration = "STAR",
    title = "{Inclusive charged hadron elliptic flow in Au + Au collisions at $\sqrt{s_{_{\text{NN}}}}$ = 7.7 - 39 GeV}",
    eprint = "1206.5528",
    archivePrefix = "arXiv",
    primaryClass = "nucl-ex",
    doi = "10.1103/PhysRevC.86.054908",
    journal = "Phys. Rev. C",
    volume = "86",
    pages = "054908",
    year = "2012"
}

@article{Allison:2016lfl,
    author = "Allison, J. and others",
    title = "{Recent developments in Geant4}",
    reportNumber = "FERMILAB-PUB-16-447-CD",
    doi = "10.1016/j.nima.2016.06.125",
    journal = "Nucl. Instrum. Meth. A",
    volume = "835",
    pages = "186--225",
    year = "2016"
}

@article{Ogul:2010si,
    author = "Ogul, R. and others",
    title = "{Isospin dependent multifragmentation of relativistic projectiles}",
    eprint = "1006.3723",
    archivePrefix = "arXiv",
    primaryClass = "nucl-ex",
    doi = "10.1103/PhysRevC.85.019903",
    journal = "Phys. Rev. C",
    volume = "83",
    pages = "024608",
    year = "2011",
    note = "[Erratum: Phys.Rev.C 85, 019903 (2012)]"
}

@article{Benecke:1969sh,
    author = "Benecke, J. and Chou, T. T. and Yang, Chen-Ning and Yen, E.",
    title = "Hypothesis of Limiting Fragmentation in High-Energy Collisions",
    doi = "10.1103/PhysRev.188.2159",
    journal = "Phys. Rev.",
    volume = "188",
    pages = "2159--2169",
    year = "1969"
}

\end{document}